\documentclass[aps,twocolumn,superscriptaddress,floatfix]{revtex4-2} 
\usepackage{color}
\usepackage{amsmath}
\usepackage{graphicx}
\usepackage{bm}
\usepackage{braket}
\usepackage{colortbl}
\usepackage{comment}
\usepackage{graphicx}
\usepackage[]{natbib}
\usepackage{hyperref}
\usepackage{url}
\usepackage{here}

\usepackage{xcolor}
\usepackage{soul}
\colorlet{soulyellow}{yellow!20}
\sethlcolor{soulyellow}
\soulregister\cite7
\soulregister\ref7
\begin{document}
\title{ Coherent interaction of a-few-electron quantum dot with a terahertz optical resonator}
\author{Kazuyuki Kuroyama}
\email[]{kuroyama@iis.u-tokyo.ac.jp}
\affiliation{Institute of Industrial Science, The University of Tokyo,
4-6-1 Komaba, Meguro-ku, Tokyo 153-8505, Japan}
\author{Jinkwan Kwoen}
\affiliation{Institute for Nano Quantum Information Electronics, The University of Tokyo,
4-6-1 Komaba, Meguro-ku, Tokyo 153-8505, Japan}
\author{Yasuhiko Arakawa}
\affiliation{Institute for Nano Quantum Information Electronics, The University of Tokyo,
4-6-1 Komaba, Meguro-ku, Tokyo 153-8505, Japan}
\author{Kazuhiko Hirakawa}
\email[]{hirakawa@iis.u-tokyo.ac.jp}
\affiliation{Institute of Industrial Science, The University of Tokyo,
4-6-1 Komaba, Meguro-ku, Tokyo 153-8505, Japan}
\affiliation{Institute for Nano Quantum Information Electronics, The University of Tokyo,
4-6-1 Komaba, Meguro-ku, Tokyo 153-8505, Japan}
\begin{abstract}
We have investigated light-matter hybrid excitations in a quantum dot (QD)-terahertz (THz) optical resonator coupled system. We fabricate a gate-defined QD in the vicinity of a THz split-ring resonator (SRR) by using a AlGaAs/GaAs two-dimensional electron system (2DES). By illuminating the system with THz radiation, the QD shows a current change whose spectrum exhibits coherent coupling between the electrons in the QD and the SRR as well as coupling between the 2DES and the SRR. The latter coupling enters the ultrastrong coupling regime and the coupling between the QD and the SRR is also very close to the ultrastrong coupling regime, despite the fact that only a few electrons reside in the QD.    
\end{abstract}

\maketitle
Coherent light-matter coupling has long been studied intensively \cite{WeisbuchPRL1992, ThompsonPRL1992} in the context of coherent manipulation of material properties such as optical non-linearity, electron transport and electronic phases \cite{KhitrovaNatPhys2006, FriskKockum2019, RevModPhys.91.025005, ParaviciniNP2019, AshidaPRX2020}. High mobility two-dimensional (2D) electron systems in AlGaAs/GaAs heterostructures combined with terahertz (THz) sub-wavelength optical resonators are one of such systems that can realize ultrastrong coupling \cite{CiutiPRB2005, TodorovPRL2010, ScalariScience2012, ParaviciniNP2019, JeanninNL2019, JeanninNL2020, MavronaACSP2021, AppuglieseScience2022}. Since the coupling strength between electrons and optical resonators scales with $\sqrt{N_e}$, where $N_e$ is the number of electrons involved in the coupled system, collective excitations of electrons have often been used to achieve ultrastrong coupling. To make a significant progress in quantum information processing based on the circuit quantum electrodynamics, it is essential to realize ultrastrong coupling in a-few-electron systems \cite{TodorovPRX2014}. However, the realization is very challenging, because collective enhancement of electron-photon interaction cannot be exploited.

Nevertheless, a few attempts have been made to realize ultrastrong coupling by using a fewer number of electrons. Previous studies demonstrated ultrastrong coupling in a subwavelength LC-resonator which contained less than one hundred electrons in a sub-$\mathrm{\mu}$m gap region \cite{KellerNanoLett2017, Rajabali2022NatComm}. Valmora et al. realized deep strong coupling between a carbon nanotube QD and a THz optical resonator and observed novel resonator-induced conductance suppression \cite{Valmorra2021}. Furthermore, Scarlino et al. demonstrated ultrastrong coupling in the microwave frequency range between a GaAs double QD and a superconducting resonator that consists of a Josephson junction array \cite{ScarlinoPRX2022}. However, the structures reported so far were rather complex or they may not be suitable for integration to realize large quantum coherent systems. Therefore, realization of hybrid quantum systems by using a simpler structure is highly desired.

In this work, we have fabricated a gate-defined QD in the vicinity of a gap of a THz split-ring resonator (SRR) located on a 2D electron system (2DES) and investigated interactions between electrons and a THz light field generated near the gap region of the SRR, by measuring the conductance through the QD. We have measured a THz-induced photocurrent through the QD under a magnetic field, $B$. Electronic excitations of 2D electrons and electrons in the QD exhibit a remarkably large anti-crossing around the SRR resonance frequency. The obtained energy dispersion of the coupled system can be understood in terms of simultaneous coupling between the 2D electrons and the SRR and that between the electrons in the QD and the SRR. The ratio of the Rabi frequency for the 2DES-SRR coupling to the SRR resonance frequency is found to be about 0.1, indicating that the system enters the ultrastrong coupling regime. Furthermore, the coupling between the QD and the SRR is also found to be very close to the ultrastrong coupling regime, despite the small number of electrons in the QD.

\begin{figure*}[t]
\centering
\includegraphics[width=\linewidth]{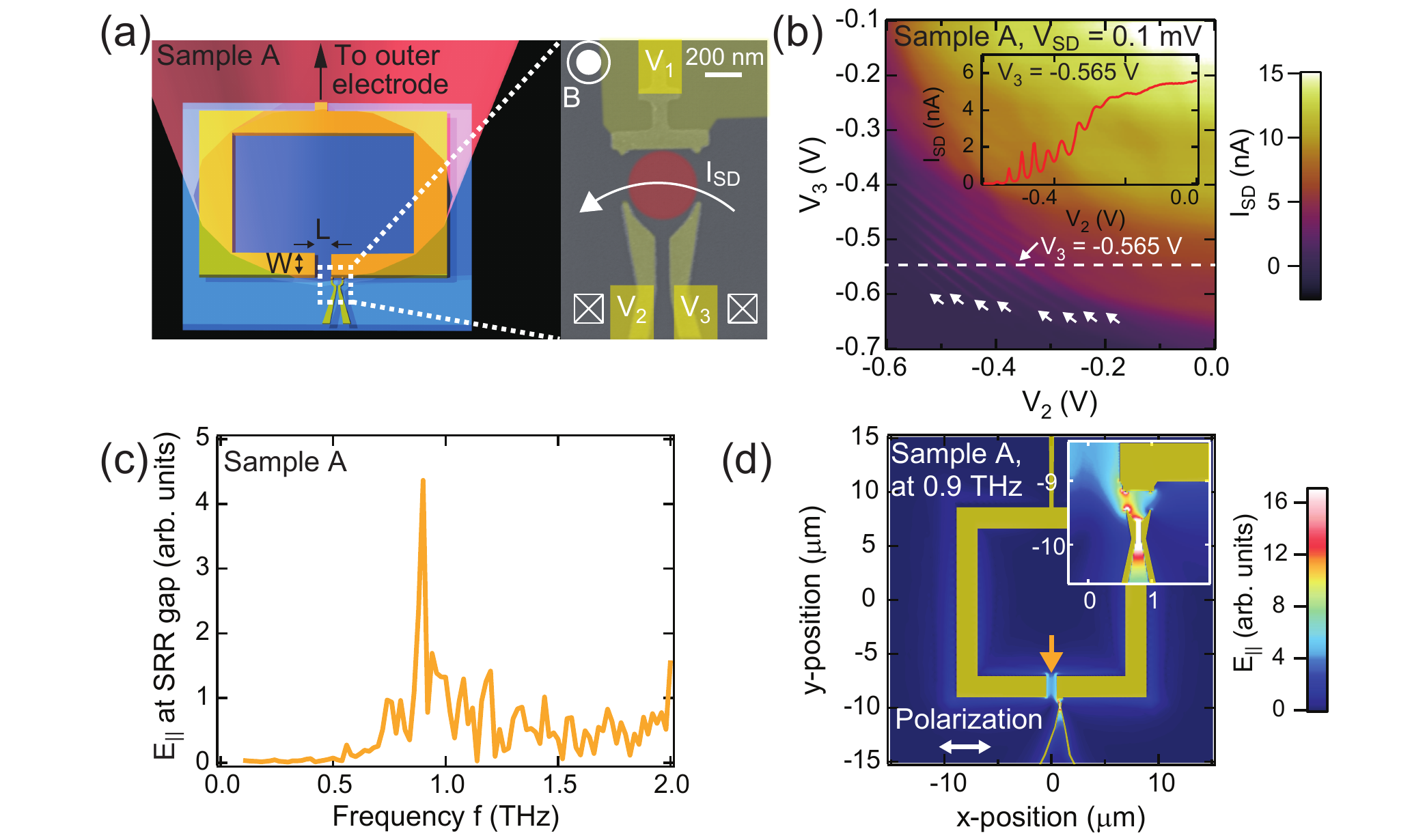} 
\caption{(a) Left panel shows an illustration of our QD-SRR coupled sample. The gap length, $L$, was designed to be 1 $\mathrm{\mu}$m. The right panel shows a scanning electron microscope image of the QD region. The white crosses represent AuGeNi ohmic contacts. (b) DC current through the QD measured at $V_{\mathrm{SD}} = 0.1$ mV as a function of the two side gate voltages. The current peaks indicated by white arrows are the Coulomb oscillation of the QD. The inset is the current measured through the QD as a function of $V_2$ along the white dashed line. (c) Spectrum of the in-plane electric field strength $E_{\parallel}$ in the SRR gap, calculated by the frequency-domain finite element method (FDFEM). The half-wavelength ($\lambda$/2)-resonance mode is located at around 0.9 THz. (d) Numerically calculated intensity map of $E_{\parallel}$ in the 2D electron layer plane ($\sim$ 100 nm below the surface) plotted at 0.9 THz of the incident radiation. The inset is a magnified view around the QD location. The THz field is strongly enhanced near the SRR gap region (orange arrow) and the QD region as well.}
\label{fig1} 
\end{figure*}

Our QD-SRR sample was fabricated by using a modulation-doped AlGaAs/GaAs high mobility heterojunction wafer. We measured two slightly different types of QDs, which were named samples A and B. The geometry of sample A is depicted in Fig. \ref{fig1}(a) \cite{HansonRevModPhys2007}. The SRR was formed by depositing a Ti/Au layer. It works as a subwavelength THz optical resonator that utilizes a concept of a lumped LC resonance circuit and has a shape of letter C, as shown on the left of the illustration \cite{SmithPRL2000}. The size of our SRR (the outer rim) was 18$\times$18 $\mathrm{\mu m^2}$. The gap width, $W$, and length, $L$, were designed to be 2 $\mathrm{\mu m}$ and 1 $\mathrm{\mu m}$, respectively. The QD was formed very near the gap of the SRR by applying appropriate negative voltages to $V_1$ and the two side gates, $V_2$ and $V_3$. Note that 2D electrons beneath the SRR are depleted, while those in the interior of the SRR ring and in the gap region remain. The source-drain bias voltage, $V_{\mathrm{SD}}$, was applied across the QD through the ohmic contacts illustrated by white crosses in Fig. \ref{fig1}(a). All the measurements were performed at a base temperature of 0.32 K of a $^3$He cryostat (refer to the entire setup illustrated in Fig. S1). The sample was attached on a Si hemispherical lens to focus the incident monochromatic THz radiation onto the sample from the substrate side. Monochromatic THz radiation was generated by the difference frequency generation, using two frequency-tunable laser diodes together with a uni-traveling carrier photodiode (UTC-PD). The output frequency can be tuned from 0.1 to 3 THz. The THz output power is approximately 1 $\mathrm{\mu W}$ at around 0.1 THz and decreases to 100 pW at 3 THz. The output power of the UTC-PD was periodically modulated at 83 Hz, and we measured the THz-induced photocurrent through the QD.

To characterize the electron transport through the QD without THz radiation, we measured the conductance at $B=0$ as a function of $V_2$ and $V_3$ when the 2D electrons beneath the SRR were depleted by applying $V_1$, as shown in Fig. \ref{fig1}(b). Coulomb oscillations are clearly observed (white arrows in the figure), indicating that a QD is formed at the designed position and the electron number can be precisely controlled.

Next, by using the frequency-domain finite element method (FDFEM), we calculated the spectrum of the in-plane electric field, $E_{\parallel}=\sqrt{|E_x|^2+|E_y|^2}$ ), in the SRR gap region. Note that we took into account the actual structure of the metal electrodes but neglected the 2D electron layer. The calculated resonance spectrum shown in Fig. \ref{fig1}(c) has a peak at 0.9 THz, which is the half-wavelength ($\lambda$/2)-resonance mode, i.e., the lowest excitation mode of the SRR. Figure \ref{fig1}(d) shows the spatial distribution of $E_{\parallel}$ at the resonance frequency in the plane 100-nm deep from the surface. The electric field enhancement is clearly observed near the SRR gap (see an orange arrow) and is extended to the region where the QD is located, thanks to the antenna effect of the side-gate electrodes. 
\begin{figure*}[t]
\centering
\includegraphics[width=\linewidth]{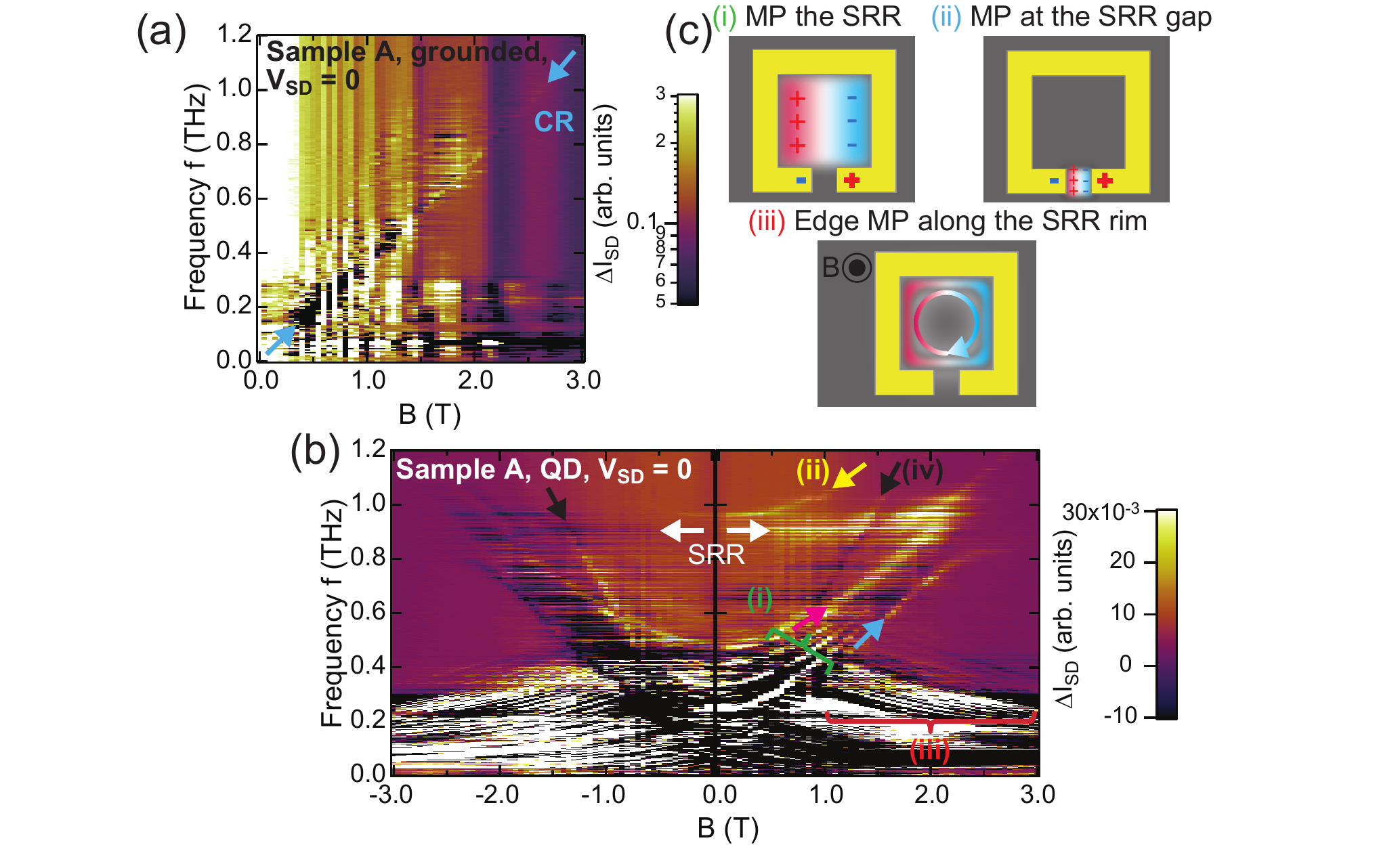} 
\caption{(a) Color-coded map of the THz-induced photocurrent measured as a function of the incident THz frequency, $f$, and the external magnetic field, $B$, at $V_{\mathrm{SD}} = 0$, when $V_2$ and $V_3$ are grounded. The cyclotron resonance signal appears, as indicated by a blue arrow. (b) Color-coded map of the THz-induced photocurrent through the QD (sample A) measured at $V_{\mathrm{SD}} = 0$ as a function of $f$ and $B$. The $B$-field independent signal at 0.9 THz is assigned to the resonant absorption by the SRR, as expected from Fig. \ref{fig1}(b). The signal which appears above $B = 1.5$ T (indicated by a blue arrow) shows an anti-crossing against the SRR resonant absorption. The numbers (i)-(iii) respectively correspond to be the modes illustrated in Fig. \ref{fig2}(c). The second harmonic of the cyclotron excitation is indicated by black arrows. (c) Pictorial explanations for the magnetoplasmon (MP) modes excited in the 2D electron system surrounded by the SRR.}
\label{fig2} 
\end{figure*}

As a reference experiment, we first measured a photocurrent as a function of the $B$-field and the frequency of the incident THz radiation,$f$, at $V_{\mathrm{SD}}=0$, while the side gates $V_2$ and $V_3$ are grounded, as shown in Fig. \ref{fig2}(a). The $B$-field-independent signals observed below 0.3 THz are induced by the lead, which is connected to the SRR for the sake of wire bonding (not shown in Fig. \ref{fig1}(a)) and works as an antenna for low frequency electromagnetic waves. The vertical modulation pattern that depends only on the $B$-field in the spectral map originates from the Shubnikov-de Haas (SdH) oscillation. A signal indicated by a blue arrow is the cyclotron resonance of the 2D electrons. Since the cyclotron resonance signal appears over a wide range between the SdH peaks, the cyclotron excitations observed in the photocurrent are likely to take place in the quantum Hall edge channels \cite{HirakawaPRB2001}. 

Next, we measured the photocurrent spectrum with the QD formed by applying voltages to the finger gates, $V_2$ and $V_3$. Figure \ref{fig2}(b) shows a color-coded photocurrent map measured at $V_{\mathrm{SD}}=0$ as a function of $B$ and $f$ (see Supplemental Material, Sec. V. for the same colormap without the indicators). Detailed experimental methods and setup can be found in Supplemental Material, Sec. I. A signal located at around 0.90 THz is assigned to the resonant absorption of the SRR, as expected from Fig. \ref{fig1}(b). By fitting a Lorentzian function to the SRR absorption peak in Fig. \ref{fig2}(b), the quality (Q)-factor of the SRR is estimated to be 13, corresponding to the full width at half maximum (FWHM) of about 68 GHz, which is similar to the one reported in the previous studies \cite{ParaviciniNP2019}. Note that several other signals which depend on the $B$-field are observed, suggesting that they originate from electronic excitations in the system. Since the electronic transport through a QD is sensitive to its electromagnetic environment, the measured QD conductance shown in Fig. \ref{fig2}(b) reflects various kinds of electronic excitations in the system. For example,
\begin{itemize}
\item[(i)] 2D magnetoplasmon excitations in the interior of the SRR (indicated in green in Fig. \ref{fig2}(b); also see (i) of Fig. \ref{fig2}(c)).
\item[(ii)] 2D magnetoplasmon excitation in the SRR gap (indicated by a yellow arrow in Fig. \ref{fig2}(b); also see (ii) of Fig. \ref{fig2}(c)).
\item[(iii)] Edge-magnetoplasmon modes (indicated in red in Fig. \ref{fig2}(b); also see (iii) of Fig. \ref{fig2}(c)).
\item[(iv)] The second harmonic of the cyclotron excitation of the 2D electrons (see a black arrow in Fig. \ref{fig2}(b)).
\end{itemize}
Although these 2D electron excitations are very intriguing, we will not go into more detail and further discussions will be made in Supplemental Material, Sec. III. Here, we focus ourselves on a fact that the conductance peak indicated by a blue arrow shows an anti-crossing behavior when it approaches the SRR resonance frequency. The observed anti-crossing indicates that the 2D electrons are coherently coupled with the nearby SRR.

\begin{figure}[t]
\centering
\includegraphics[width=\linewidth]{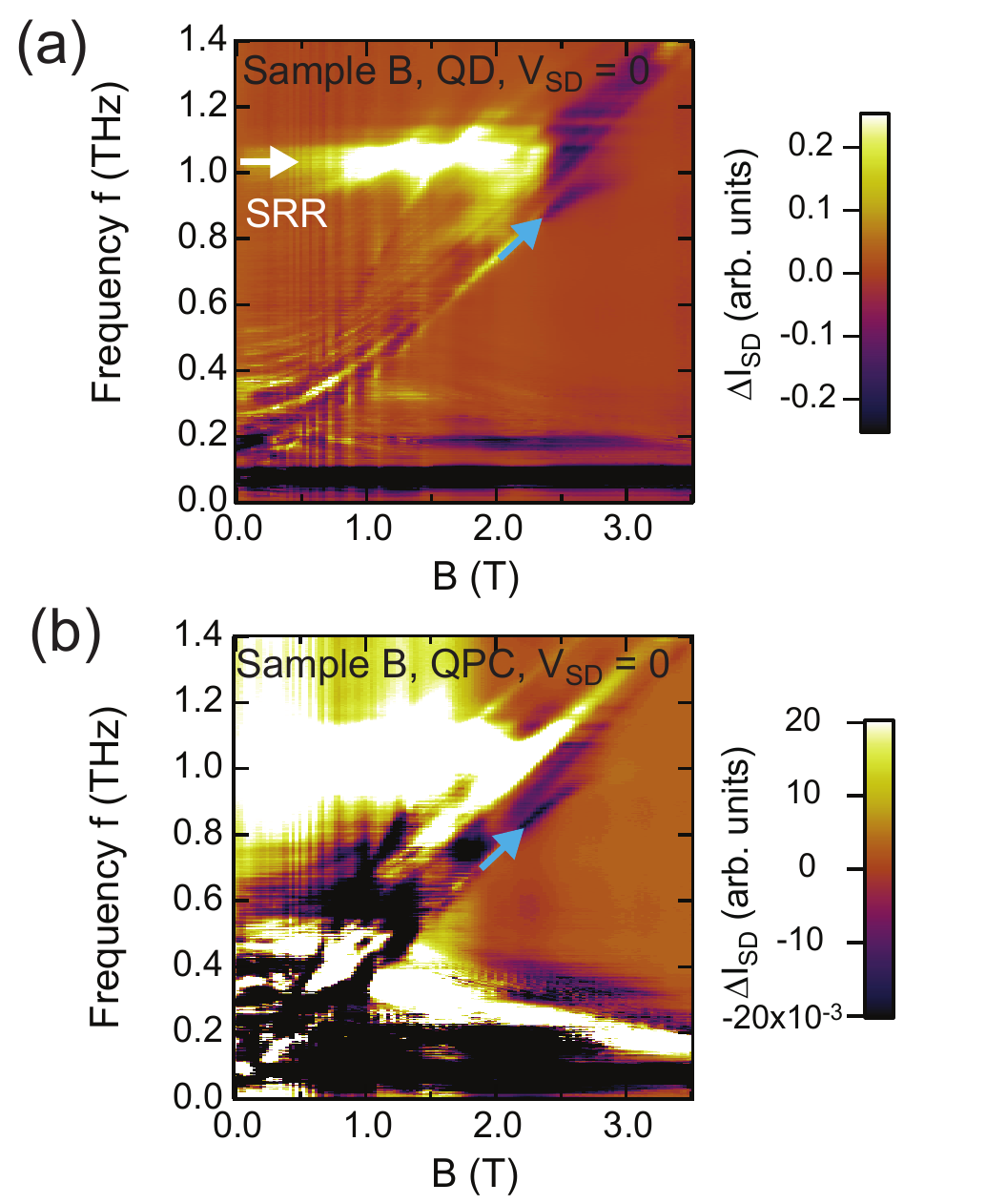} 
\caption{(a) Color-coded map of the THz-induced photocurrent through the QD (sample B) measured at $V_{\mathrm{SD}}=0$ as a function of $f$ and $B$. The anti-crossing similar to that in Fig. \ref{fig2}(b) is observed (blue arrow). (b) Color-coded map of the THz-induced photocurrent through the QPC (sample B) measured at $V_{\mathrm{SD}}=0$ as a function of $f$ and $B$ . The similar anti-crossing is observed (see the region indicated by a blue arrow).}
\label{fig3} 
\end{figure}

Next, we discuss electrons in which part generate the anti-crossing behavior against the SRR. To do so, let us compare between the THz-induced photocurrent map when the QD is formed and that when a quantum point contact (QPC) is formed by changing the gate configuration. Figures \ref{fig3}(a) and \ref{fig3}(b) are the color-coded maps of the THz-induced photocurrent measured for the QD and QPC configurations, respectively. Note that these spectra were measured on another sample (sample B) (see Supplemental Material, Sec. II for the sample design). The QPC was formed in the vicinity of the gap of the SRR by applying only $V_2$ and leaving $V_3$, $V_{\mathrm{p}}$ (the plunger gate voltage) grounded, as illustrated in Fig. S3 of Supplemental Material, Sec. II. When Figs. \ref{fig3}(a) and \ref{fig3}(b) are compared, the features in the photocurrent map for the QPC (Fig. \ref{fig3}(b)) are almost identical to those observed for the QD (Fig. \ref{fig3}(a)), including the distinct anti-crossing behavior (see blue arrows). Therefore, the results suggest that the anti-crossing behavior mainly originates from the cyclotron excitations of the 2D electrons near the SRR. The dependence on the incident THz polarization is shown in Supplemental Material, Sec. V.

\begin{figure}[t]
\centering
\includegraphics[width=\linewidth]{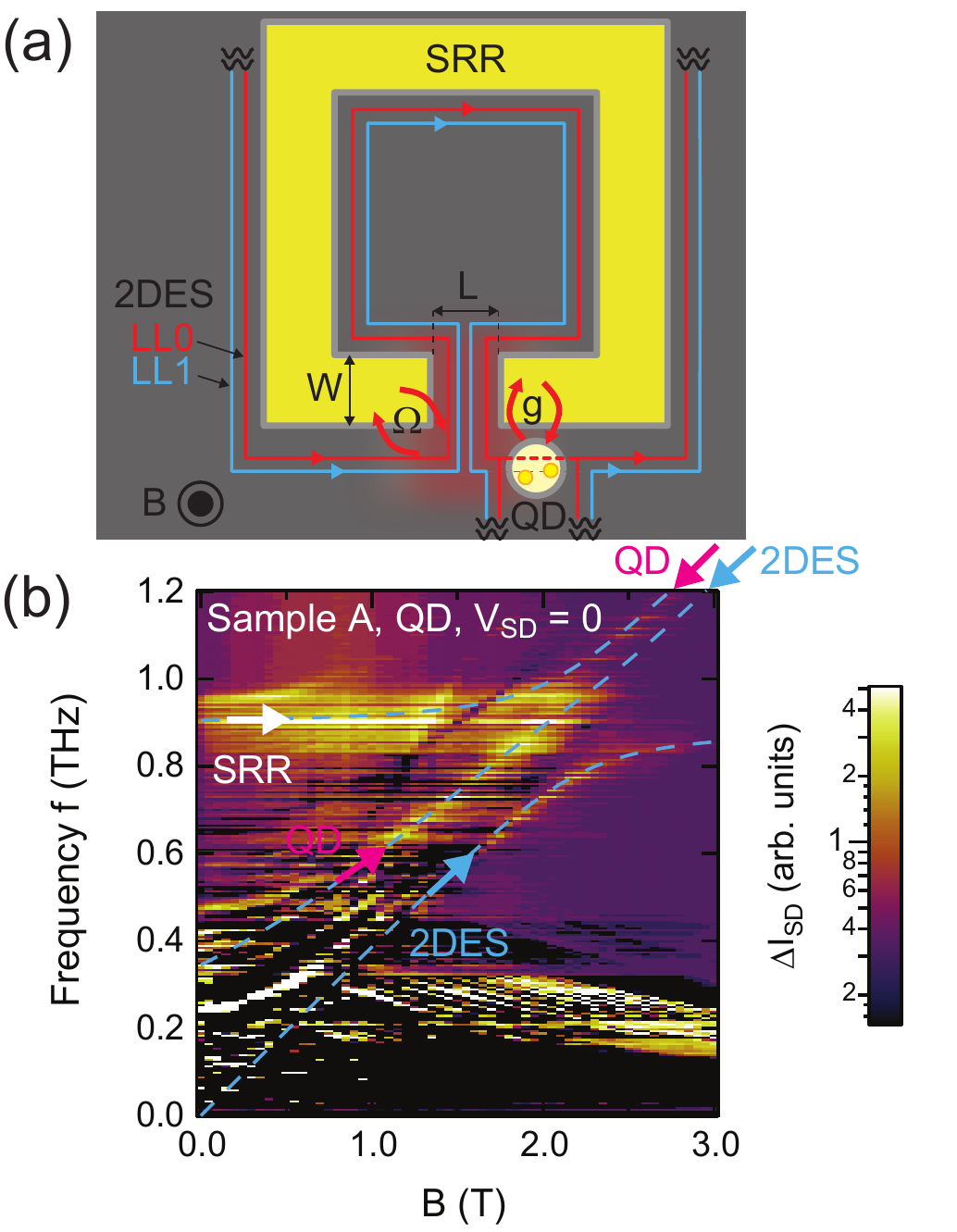} 
\caption{(a) Illustration of the 2DES-SRR-QD coupled system considered in the present theoretical model. The 2D electrons which are located near the SRR gap and are coupled to the SRR are represented by a red region. Red and blue lines pictorially represent the quantum Hall edge channels propagating along the SRR when the filling factor of the bulk region is assumed to be 2. ``LL0'' and ``LL1'' are the lowest and second lowest Landau levels, respectively. (b) Calculated energy dispersions of the 2DES-SRR-QD coupled system, which are plotted by blue dashed lines on the color-coded map of the THz-induced photocurrent through the QD.}
\label{fig4} 
\end{figure}

Furthermore, we have found that most of the observed signals in the photocurrent map shown in Fig. \ref{fig2}(b) change their polarities when the $B$-field direction is reversed, indicating that the signals originate from the Hall conductance of the 2D electrons. Therefore, electrons in the chiral quantum Hall edge states propagating near the SRR, which are represented by blue and red lines in Fig. \ref{fig4}(a) for the case of the filling factor $\nu=2$ (the spin degeneracy neglected), are very likely to create the anti-crossing with the SRR. Note that ``LL0'' and ``LL1'' in Fig. \ref{fig4}(a) are the lowest and second lowest Landau levels, respectively.

Let us estimate the coupling strength of the light-matter interaction in this system. So far, we have discussed the coupling between the 2D electrons in the quantum Hall edge channel and the SRR. However, electrons confined in the QD or QPC should also interact with the SRR, since they are placed near the SRR gap, where the THz field is strongly enhanced (see Fig. \ref{fig1}(d)). Indeed, one more spectral line is discernible, which is denoted by a magenta arrow in Fig. \ref{fig2}(b). It is very likely that this line represents the resonant excitation in the QD (see Supplemental Material, Sec. IV. for more supporting data). Therefore, the actual coupled system that we measure can be illustrated as Fig. \ref{fig4}(a). We calculate the energy dispersions of the 2DES-SRR-QD coupled system, which are derived from the following Hamiltonian that describes the excitations of $\ket{e_{\mathrm{2DES}},n=0,g_{\mathrm{QD}}}$, $\ket{g_{\mathrm{2DES}},n=1,g_{\mathrm{QD}}}$, and $\ket{g_{\mathrm{2DES}},n=0,e_{\mathrm{QD}}}$, where $g_i$ and $e_i$ ($i=$2DES,QD) are the ground and excited states of the 2DES and QD, and $n$ is the photon number in the SRR. 
\begin{align}
\mathcal{H}=\hbar
\left(
  \begin{array}{ccc}
    \omega_c(B) & \Omega(B) & 0 \\
    \Omega(B) & \omega_{\mathrm{SRR}} & g(B) \\
    0 & g(B) & \omega_{\mathrm{QD}}(B)
    \label{eq:Hamiltonian}
  \end{array}
\right).
\end{align}
Note that we do not take into account the squared terms of the vector potential. $\omega_c= eB/m^*$ and $\omega_{\mathrm{QD}}=\sqrt{(\omega_0^2+(\omega_c/2)^2)}+\omega_c/2$ are the cyclotron frequency and the frequency of electrons in the QD, respectively \cite{FockZP1928, DarwinCUP1931}. Here, $\hbar\omega_0$ is the QD orbital spacing energy at $B = 0$, and we used $m^*=0.071m_0$ as the electron effective mass in GaAs. $\omega_0/2\pi$ was determined to be 345 GHz from the fitting. $g$ and $\Omega$ are the Rabi frequency between the electrons in the QD and the SRR and that between the 2DES and the SRR, respectively. Their $B$-field dependences are $g\simeq g_0\sqrt{B/B_{\mathrm{resQD}}}$ obtained from the Fock-Darwin wavefunctions and $\Omega=\Omega_0\sqrt{B/B_{\mathrm{res2DES}}}$, where $B_{\mathrm{resQD}}$ and $B_{\mathrm{res2DES}}$ are defined by $\omega_{\mathrm{QD}}(B_{\mathrm{resQD}})\equiv \omega_{\mathrm{SRR}}$, and $\omega_c(B_{\mathrm{res2DES}})\equiv \omega_{\mathrm{SRR}}$ \cite{BartoloPRB2018, CiutiPRB2021}. We derived the polariton energies by diagonalizing the Hamiltonian matrix. The fitted energy dispersion curves with $\omega_{\mathrm{SRR}}/2\pi=0.905$ THz are shown in Fig. \ref{fig4}(b) by blue dashed lines. The calculated lines reproduce the spectral features very well, showing that the signal marked with a magenta arrow is very likely to be the electron excitation in the QD, although the QD signal cannot be clearly seen due to the overlap with MP signals for $f < 0.6$ THz. From this analysis, we determine the Rabi frequency of the 2DES-SRR coupling $\Omega_0/2\pi$ to be $\sim 100$ GHz and that of the QD-SRR coupling $g_0/2\pi$ to be $\sim 50$ GHz. $\Omega_0/\omega_{\mathrm{SRR}}$ is as large as 0.1, meaning that the 2DES-SRR coupled system is in the ultrastrong coupling regime. Furthermore, the QD-SRR coupling is also very close to the ultrastrong coupling regime, despite the fact that the QD accommodates only a few electrons inside. This can be understood by strong THz field enhancement at the QD position by the side gates and the SRR.

In recent works, ultrastrong coupling was realized by utilizing collective excitations of AlGaAs/GaAs 2D electron systems of high carrier densities or of multiple electron layers to enhance the coupling strength. In this sense, we are going in the opposite direction, since the QD is operated in a few-electron regime. To examine whether even a single electron in a QD can have a coupling strength large enough to reach the ultrastrong coupling regime, we have made a simple estimation of coupling strength between a single electron dipole in a QD and the SRR. The electric field created by a single photon of an energy $\hbar\omega_{\mathrm{SRR}}$ in the SRR gap (see Figs. \ref{fig1}(a) and \ref{fig4}(a)) is expressed as $E_{\mathrm{gap}}\simeq \sqrt{\hbar\omega_{\mathrm{SRR}}/\epsilon WL^2}$, where $\epsilon=[12.89r+1(1-r)] \epsilon_0$ is the effective dielectric constant for a GaAs-vacuum interface, and $r$ was set to be 0.3 by fitting analysis of the magnetoplasmon excitations as discussed in Supplemental Material. Sec. III. For simplicity, we assumed that the electric field extends perpendicular to the SRR by a distance $\simeq L$ \cite{ParaviciniPRB2017}. Note that, according to the numerical calculation shown in Fig. \ref{fig1}(b), the electric field around the QD position is approximately 4 times larger than that in the gap region. The QD confinement length of a single electron in the QD under perpendicular $B$-field is described as $l=l_{B} [1/4+(\omega_c/\omega_{\mathrm{QD}})^2 ]^{1/4}$, where $l_B=\sqrt{\hbar/eB}$ is the magnetic length \cite{PhysRevB.84.235309}. Since the QD-SRR coupling is realized by the electric-dipole interaction, the Rabi frequency of a single electron in the QD is calculated as $g_0\simeq el(4E_{\mathrm{gap}})/\hbar$. Using this relationship for our sample geometry at $B_{\mathrm{resQD}} \sim 1.96$ T, the Rabi frequency for a single electron and a single photon is calculated to be $g_0/2\pi \sim 58.9$ GHz, which is close to the experimentally estimated $g_0$. We can trace the origin of the very strong coupling between a single electron in the QD and the SRR back to a large enhancement in the electric field due to the small and sharp edges of the finger gates.

In conclusion, we have investigated THz-induced photocurrent through the QD in the 2DES-SRR-QD coupled system. We have observed a remarkably large anti-crossings generated by the coupling between the SRR and electronic excitations in the QD as well as 2DES. The calculated energy dispersion of the 2DES-SRR-QD polaritonic states is in good agreement with the observed THz photocurrent spectrum. The normalized coupling strength $\Omega_0/\omega_{\mathrm{SRR}}$ is as large as 0.1, indicating that the 2DES-SRR coupled system is in the ultrastrong coupling regime. The $B$-field polarity dependence of the photocurrent signal indicates that the 2D electrons that strongly interact with the SRR lie in the quantum Hall edge states along the edge of the SRR. Furthermore, the electronic excitations in the QD also exhibit coherent coupling with the SRR, which is very close to the ultrastrong coupling regime, thanks to the strong enhancement of THz electric fields by the antenna effect of the side gates. The QD-SRR coupling would be better resolved by suppressing the radiation losses in the split-ring resonator \cite{LiPRB2009}. The present ultrastrongly coupled system allows us to investigate quantum transport of polaritons by using electrically-controllable quantum nanostructures. These studies greatly benefit the understanding of fundamental physics of polaritons and their manybody effects in the polaritonic quantum phase transition \cite{AshidaPRX2020, TodorovPRX2014}. Furthermore, owing to its simple design, the present QD-SRR hybrid system is suitable for larger system integration.

We thank S. Komiyama, S. Iwamoto, Y. Tokura, and S. Q. Du for fruitful discussions. This work has been supported by JSPS Research Fellowship for Young Scientists (JP19J01737), and KAKENHI from JSPS (JP20K14384, JP20H05660, JP17H01038, JP15H05868, JP17H06119, JP20K15260, JP20H05218, and JP15H05700)), and JST, PRESTO Grant Number JPMJPR2255, Japan.

\global\long\def\bibsection{}%
\textbf{References} 
\bibliographystyle{unsrt}

\end{document}


\title{Supplemental Material for ``Coherent interaction of a-few-electron quantum dot with a terahertz optical resonator''}
\author{Kazuyuki Kuroyama}
\email[]{kuroyama@iis.u-tokyo.ac.jp}
\affiliation{Institute of Industrial Science, The University of Tokyo,
4-6-1 Komaba, Meguro-ku, Tokyo 153-8505, Japan}
\author{Jinkwan Kowen}
\affiliation{Institute for Nano Quantum Information Electronics, The University of Tokyo,
4-6-1 Komaba, Meguro-ku, Tokyo 153-8505, Japan}
\author{Yasuhiko Arakawa}
\affiliation{Institute for Nano Quantum Information Electronics, The University of Tokyo,
4-6-1 Komaba, Meguro-ku, Tokyo 153-8505, Japan}
\author{Kazuhiko Hirakawa}
\email[]{hirakawa@iis.u-tokyo.ac.jp}
\affiliation{Institute of Industrial Science, The University of Tokyo,
4-6-1 Komaba, Meguro-ku, Tokyo 153-8505, Japan}
\affiliation{Institute for Nano Quantum Information Electronics, The University of Tokyo,
4-6-1 Komaba, Meguro-ku, Tokyo 153-8505, Japan}

\maketitle

\section{Experimental setup}
\begin{figure}[h]
\centering
\includegraphics[width=\linewidth]{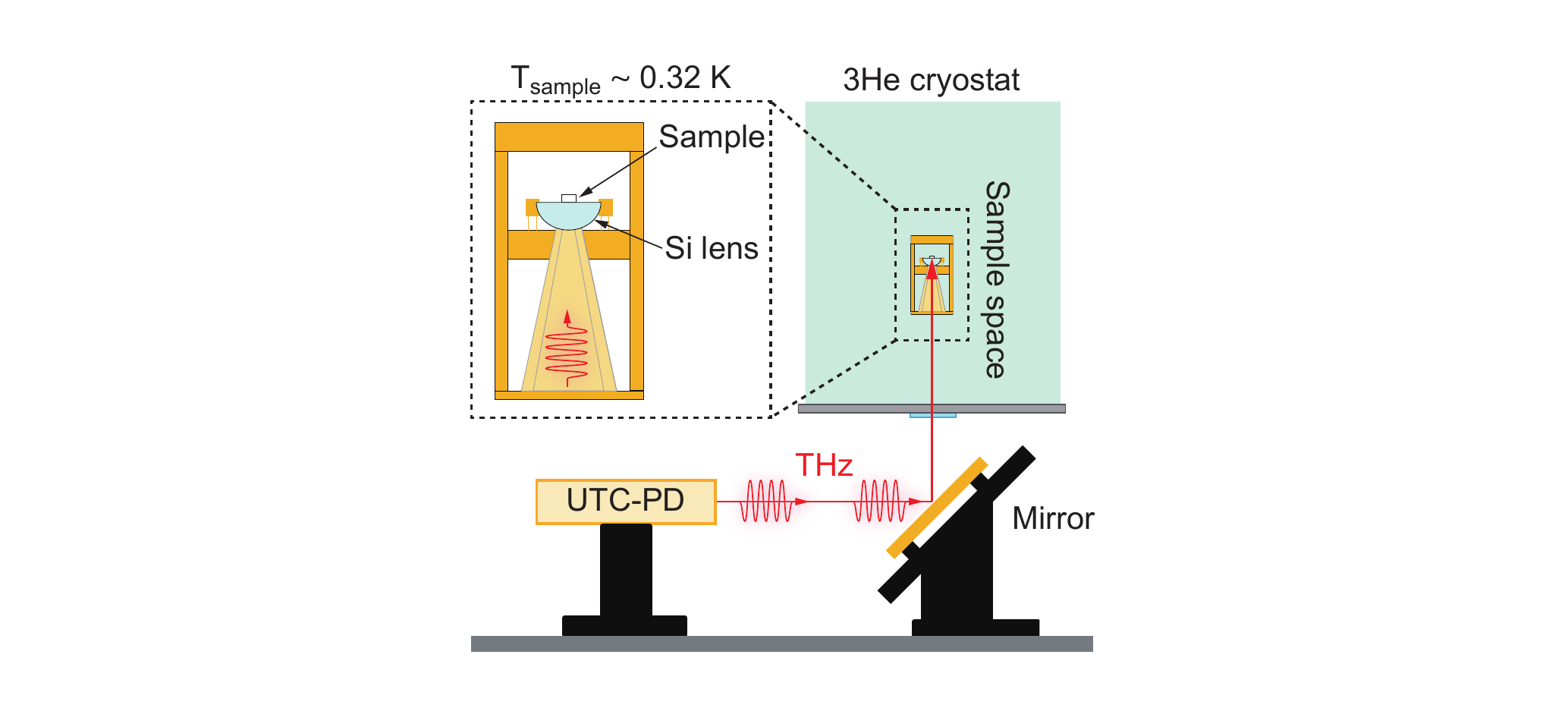} 
\caption{Optical setup for the THz spectroscopy on the QD-SRR coupled sample. Our $^3$He cryogen-free cryostat has THz optical windows on the bottom to introduce THz radiation. The output of the ultrafast, high-power photodiode (UTC-PD) is guided into the cryostat by a mirror and then focused onto the sample by using a hemispherical Si lens. To suppress the background radiation, low-pass filters were set in the cryostat.}
\label{figS1} 
\end{figure}

In our selectively doped $\mathrm{Al_{0.3}Ga_{0.7}As}$/GaAs/$\mathrm{Al_{0.3}Ga_{0.7}As}$ As double heterostructure, 2D electrons accumulate in the 20-nm-thick GaAs quantum well layer located approximately 100-nm below the surface. The mobility and carrier density determined at 0.30 K by Hall measurements were $\sim7.53\times10^5\ \mathrm{cm^2Vs}$ and $\sim1.86\times10^{11}\ \mathrm{cm^{-2}}$, respectively. The schematic of the entire experimental setup is shown in Fig. \ref{figS1}.

\section{Structure of Sample B}
\begin{figure}[h]
\centering
\includegraphics[width=\linewidth]{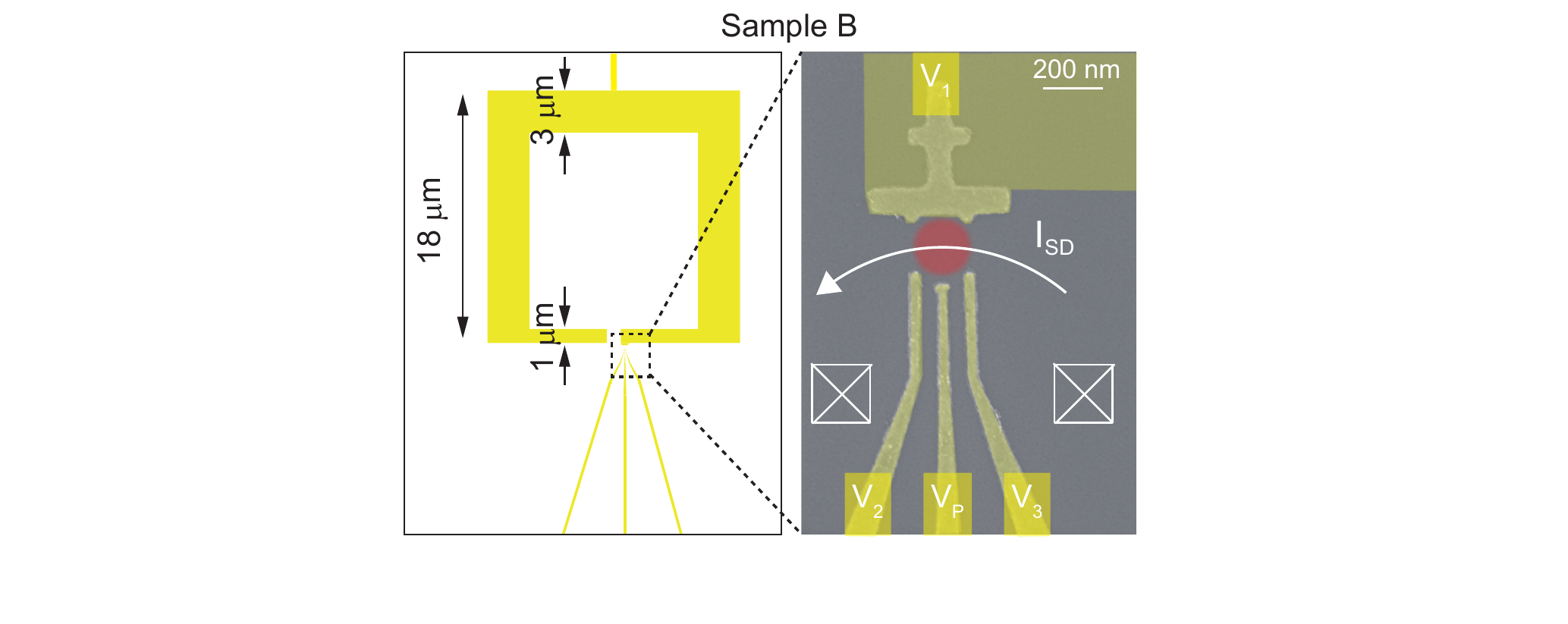} 
\caption{Left and right figures are an illustration and a scanning electron microscope image of our QD-SRR coupled sample (sample B). The design of the SRR is slightly different from that of sample A. Furthermore, we added one more side gate, i.e., the plunger gate ($V_{\mathrm{P}}$), to improve the controllability of the electrostatic potential in the QD.}
\label{figS2} 
\end{figure}

Figure \ref{figS2} shows the structure of another QD-SRR sample (sample B) used for the measurements shown in Fig. 3 of the main text. The structure is very similar to that of sample A, and we explain only the difference in the following. Concerning the SRR of sample B, the dimensions of the outer and inner rims are the same as those of sample A (18 $\mathrm{\mu m}\times$18 $\mathrm{\mu m}$, and 14 $\mathrm{\mu m}\times$ 14 $\mathrm{\mu m}$). The only difference is that the inner rim is shifted towards the SRR gap by 1 $\mathrm{\mu m}$. This difference results in a higher resonance frequency ($\sim1.05$ THz) than that of sample A ($\sim0.9$ THz). Furthermore, we added one more side gate to control the electrochemical potential in the QD, while controlling the tunnel coupling between the QD and the source-drain electrodes independently. 
 
\section{Magnetoplasmon excitations in Fig. 2}
We discuss the origin of the THz-induced signals other than the anti-crossing behavior (refer to Figs. 2(b) and 2(c)).
\begin{enumerate}
\item[(i)] 	The signals indicated in green, which shift to higher frequencies as $B$ increases, are assigned to the magnetoplasmon excitations of electrons inside the ring of the SRR (see the left top illustration (i)) of Fig. 2(c)). Similar magnetoplasmon excitations were observed in absorption experiments performed in the gigahertz and THz frequency ranges \cite{GrigelionisAPPA2012, YuOE18}. The energies of the magnetoplasmon can be calculated by the following formula.
\begin{align}
\omega_{\mathrm{p},N}=\sqrt{\frac{\pi N n_{\mathrm{2D}}e^2}{2m^*\epsilon W_{\mathrm{2D}}}} \label{eq:S1}
\end{align}
where $n_{\mathrm{2D}}$ is the electron density of the 2DES, and $W_{\mathrm{2D}}=14$ $\mathrm{\mu m}$ is the width of the 2D electron system (2DES) remaining in the SRR interior. $\epsilon=[12.89r+1(1-r)] \epsilon_0$ is an effective dielectric constant for the interface between GaAs and vacuum. Here, $r$ is a fitting parameter, whose value ranges from 0 to 1. $N$ is the harmonic number of the magnetoplasmon modes. Under a $B$-field, the magnetoplasmon energy is expressed as,
\begin{align}
\omega_{\mathrm{MP},N}=\sqrt{\omega_{\mathrm{p},N}^2+\omega_c^2}\label{eq:S2}
\end{align}
Fig. \ref{figS3}(b) is a color-coded photocurrent map of the quantum point contact (QPC) formed in sample A by using only the gate voltages $V_1$ and $V_2$. The photocurrent was measured at $V_{\mathrm{SD}}=0$ as a function of the $B$-field and the frequency of the incident THz radiation, $f$. The calculated plasmon dispersions for $N=1\sim5$ and $r=0.3$ are plotted by green curves and are in good agreement with the observed signals, as shown in Fig. \ref{figS3}(a). However, according to previous theoretical studies \cite{MikhailovPRB2005,MikhailovPRB2006}, a uniform incident electromagnetic field can excite only odd-order magnetoplasmons in a 2DES of a spatially uniform charge density. Therefore, to explain the magnetoplasmons of the even order harmonics, we need to consider non-uniformities of either the incident electromagnetic field or the charge density distribution in the 2DES. The details will be discussed in a future publication.
\begin{figure}
\includegraphics[width=\linewidth]{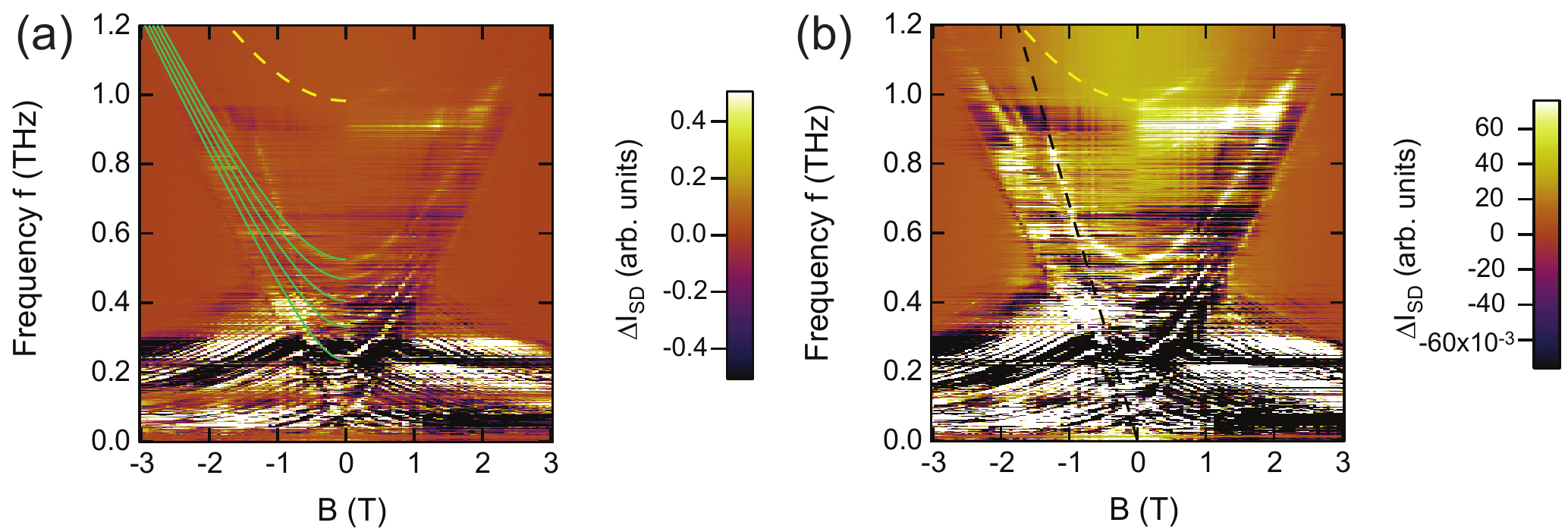} 
\caption{(a) Energy dispersions of the plasmonic excitations in the interior of the SRR (green curves) and those in the SRR gap (yellow dashed curves), fitted to the photocurrent spectrum of the QPC formed on sample A. Note that Figs. \ref{figS3}(a) and \ref{figS3}(b) are the same spectrum but plotted with the different color scale for the better visibilities of the respective plasmonic excitations.}
\label{figS3} 
\end{figure}

\item[(ii)] Magnetoplasmon excitation of 2D electrons that remain in the gap region of the SRR (see the bottom middle illustration (ii) of Fig. 2(c)) is observed slightly above the resonance frequency of the SRR (see a yellow arrow in Fig. 2(b)). The frequency of this higher energy magnetoplasmon is also in good agreement with our numerical calculation using Eqs. \ref{eq:S1} and \ref{eq:S2} for $W_{\mathrm{2D}}=0.825$ $\mathrm{\mu m}$, as plotted by a yellow curve in the right figure of Fig. \ref{figS3}(b). Although $W_{\mathrm{2D}}$ is slightly smaller than 1 $\mathrm{\mu m}$, it is reasonable because the SRR is 100-nm above the 2DES layer. 

\item[(iii)] The signals showing red shifts with increasing $B$ are also observed (red curly bracket in Fig. 2(b)). These excitations are explained by the circulating motion of the magnetoplasmons propagating along the inner edge of the SRR (see the illustration (iii) of Fig. 2(c)). Because the propagating orientation of the edge magnetoplasmon along the inner edge is opposite to that of the cyclotron motions, only the signals with the red shifts are observed. Their energy spacing is approximately 10 GHz, which is much smaller than the magnetoplasmon modes discussed above. A similar energy dispersion of the magnetoplasmon excitation was observed in a rectangle 2DES \cite{MastPRL1985} and explained by the edge magnetoplasmon excitation. However, unlike the previous study \cite{MastPRL1985}, many higher order harmonic excitations appear in our spectrum, which calls for future study.

\item[(iv)] We also note that the second harmonic of the cyclotron resonance is clearly observed in the spectrum, which is forbidden in a uniform 2D electron system by the optical selection rule. A black dashed line in Fig. \ref{figS3}(b) is the energy dispersion of the second harmonic of the cyclotron resonance calculated with the electron effective mass $m^*=0.082m_0$. When the actual electrostatic confinement felt by electrons is not parabolic, for example, due to the gate confinement electric fields, the strict selection rule may be relaxed and the higher harmonics of the cyclotron resonance may become visible. However, we do not have a clear explanation at present.
\end{enumerate}
As explained above, we have observed various kinds of magnetoplasmon excitations of 2D electrons by the photocurrent spectroscopy of the QD. They are indeed interesting but are not in the scope of the present paper. Therefore, we will not go into detail and discuss them in future publications.

\section{Comparison with the two-state Hopfield energy dispersion}
\begin{figure}
\includegraphics[width=\linewidth]{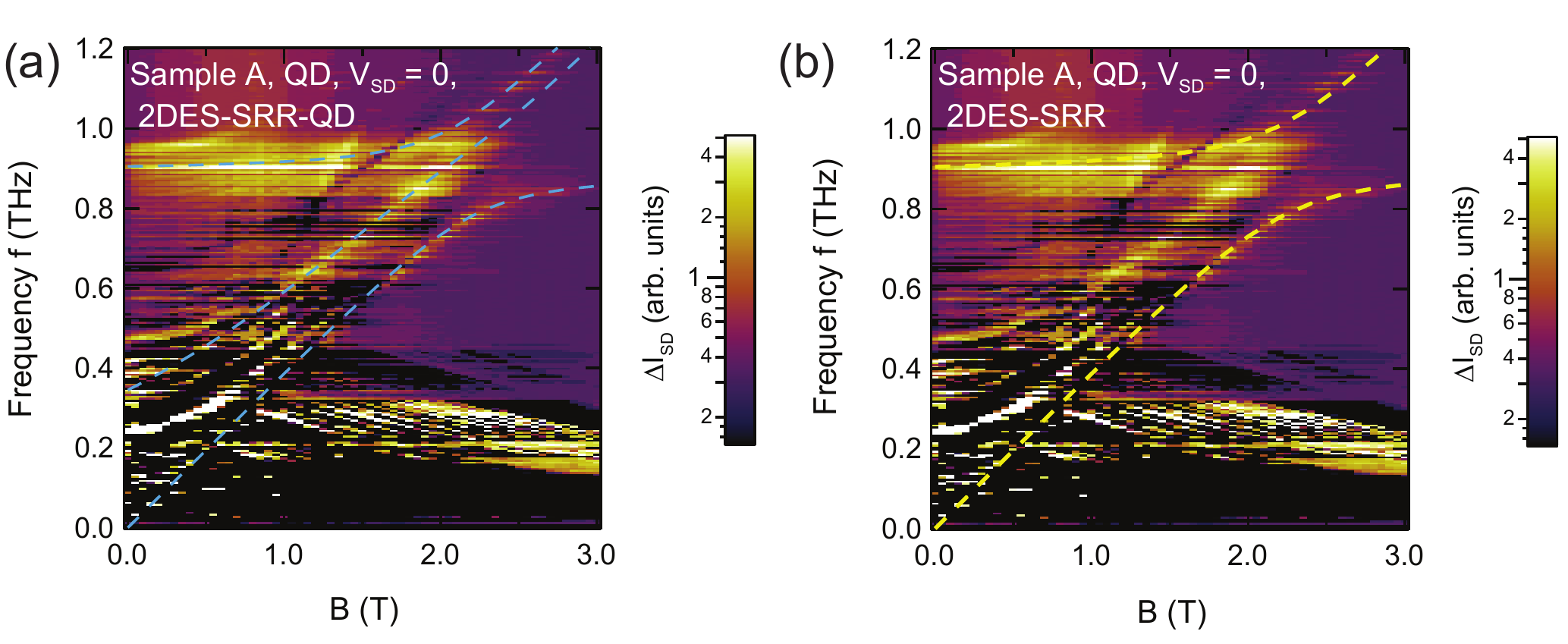} 
\caption{Comparison of the polaritonic energy dispersions between the 2DES-SRR-QD coupled system explained in the main text (a) and the 2DES-SRR coupled system explained in this section (b). Note that Fig. \ref{figS4}(a) is the same figure as Fig. 2(b) of the main text.}
\label{figS4} 
\end{figure}

In comparison with the 2DES-SRR-QD energy dispersions, we calculated the energy dispersion of the 2DES-SRR polaritons, which is derived from the full Hopfield Hamiltonian \cite{ParaviciniPRB2017, MuellerNature2020}. We try to fit the following energy dispersion to the measured spectrum by adjusting the coupling strength.
\begin{align}
\omega_{\mathrm{LP}}^{\mathrm{UP}}&=\frac{1}{\sqrt{2}}\sqrt{\omega_{\mathrm{QD}}^2+4\Omega^2+\omega_{\mathrm{cav}}^2\pm G},\\
G&=\sqrt{-4\omega_{\mathrm{QD}}^2\omega_{\mathrm{cav}}^2+(\omega_{\mathrm{QD}}^2+4\Omega^2+\omega_{\mathrm{cav}}^2)^2},
\end{align} 
where $\omega_c=eB/m^*$ is the cyclotron frequency of 2D electrons, and $m^*=0.071m_0$ is the electron effective mass in GaAs. $\Omega=\Omega_0\sqrt{B/B_{\mathrm{res2DES}}}$ is the Rabi frequency between the 2D electrons and the SRR, where $B_{\mathrm{res2DES}}\equiv m^*\omega_{\mathrm{SRR}}/e$. The Rabi frequency of the polaritons is set to be approximately $\Omega_0/2\pi =115$ GHz. The fitted curves of $\omega_{\mathrm{LP}}^{\mathrm{UP}}/2\pi$ for $\omega_{\mathrm{SRR}}/2\pi=0.905$ THz are shown in Fig. \ref{figS4}(b) by yellow dashed curves. The calculated curves reproduce the lower branch of the polariton, but the upper branch is not reproduced very well. Therefore, we conclude that the energy dispersions that include not only the SRR-2DES coupling but also the SRR-QD coupling are necessary to interpret the spectrum shown in Fig. 2(b).

\section{Polarization dependence measured on Sample B}
\begin{figure}
\includegraphics[width=\linewidth]{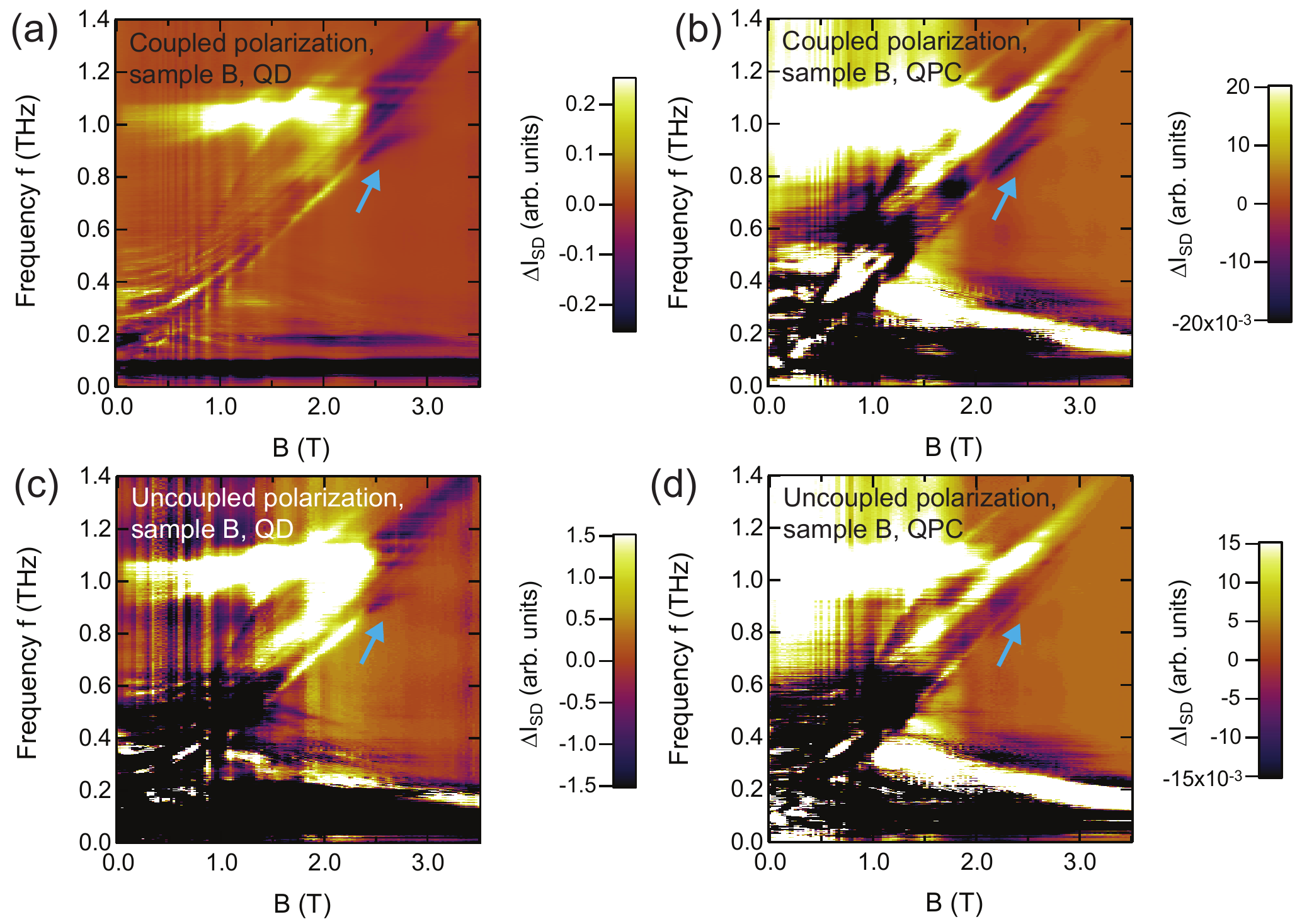} 
\caption{(a,c) Photocurrent spectra measured on sample B with the QD configuration for the coupled and uncoupled polarization conditions. Anti-crossing appears for both polarizations. (b,d) Photocurrent spectra measured on sample B with the QPC configuration for the coupled and uncoupled polarization conditions. Anti-crossing appears only when the incident polarization is directly coupled to the SRR. Note that Figs. \ref{figS5}(a) and \ref{figS5}(b) are the same figures as Figs. 3(a) and 3(b) of the main text.}
\label{figS5}
\end{figure}

We measured the incident polarization dependence of the photocurrent spectra for sample B. Figures \ref{figS5}(a) and \ref{figS5}(c) show the spectra measured for the incident polarizations perpendicular (coupled polarization) and parallel (uncoupled polarization) to the SRR gap, respectively, when the QD is formed. Their spectra are almost identical except for the intensities of the signals.

Next, we measured the photocurrent spectra for the QPC configuration for the two polarizations. The obtained spectra are plotted in Figs. \ref{figS5}(b) and \ref{figS5}(d). It should be noted that the anti-crossing indicated by a blue arrow in Figs. \ref{figS5}(b) disappears when the incident polarization is uncoupled to the SRR. 

\begin{figure}
\includegraphics[width=\linewidth]{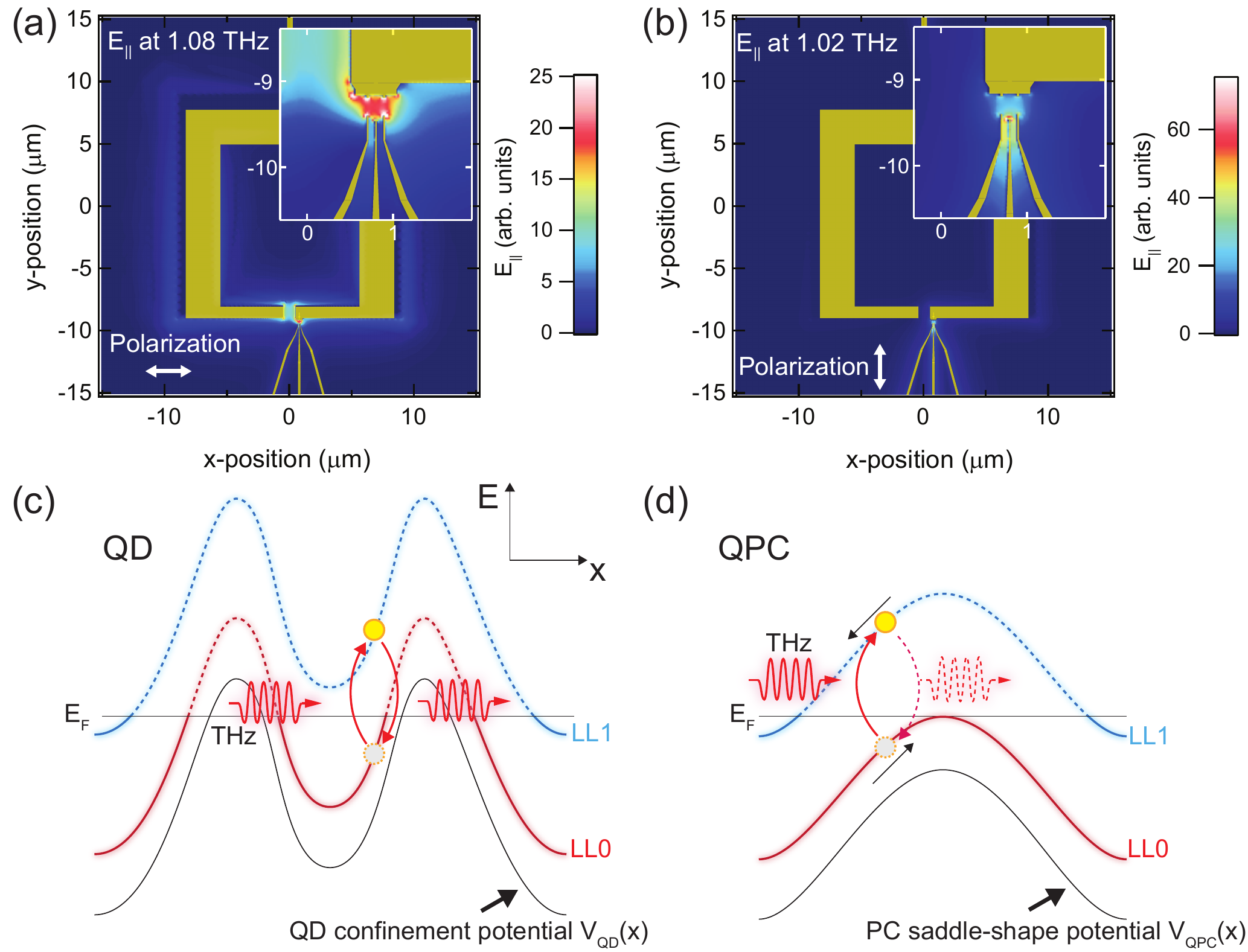} 
\caption{(a,b) Numerically calculated intensity map of the in-plane THz electric field $E_{\parallel}$ on sample B at the 2D electron layer level ($\sim 100$ nm below the surface) for the coupled and uncoupled polarization cases. The insets are a magnified view around the QD location. The THz field is not enhanced in the SRR gap region but only in the QD region. (c,d) Schematic explanations for the difference in the anti-crossing behavior between the QD and the QPC, respectively. Black dashed curves are the potential landscape in the direction of the electron propagation. ``LL0'' and ``LL1'' denote the lowest and the second lowest Landau levels, respectively (the spin degeneracy is neglected).}
\label{figS6}
\end{figure}

To understand the difference in the anti-crossing behavior between the QD and the QPC for the uncoupled THz polarization, we first consider the THz electric field distribution around the SRR and the QD. The calculated in-plane electric field $E_{\parallel}$ under THz radiation with the uncoupled polarization is shown in Fig. S6(a). Note that the incident THz frequency is set to be around 1.05 THz. Unlike the coupled incident polarization, the significant electric field enhancement appears only where the QD and the QPC are formed and does not appear in the SRR gap. Therefore, the external THz radiation does not directly excite the SRR but more likely to excite electrons in the QD and the QPC. Considering this result, we propose the following scenario: For the case of the QD, electrons in the QD can be excited between its orbitals by external THz radiation and relax by releasing photons to the SRR, as shown in an illustration of Fig. \ref{figS6}(b). Therefore, the QD can transfer the electromagnetic energy to the SRR and the coherent energy exchange between the 2DES and the SRR takes place.

In contrast, although electrons in the QPC can be excited by external radiation as well, an electron-hole separation quickly takes place because of a saddle-shape potential in a QPC. Because of this electron-hole spatial separation, the photoexcited electrons cannot relax to the original state (see Fig. \ref{figS5}(c)) and cannot transfer the photon energy back to the SRR. This difference between the QD and the QPC explains the difference in the anti-crossing behavior.

\section{Fig. 2(b) without eye guides and arrows}
\begin{figure}[h]
\includegraphics[width=\linewidth]{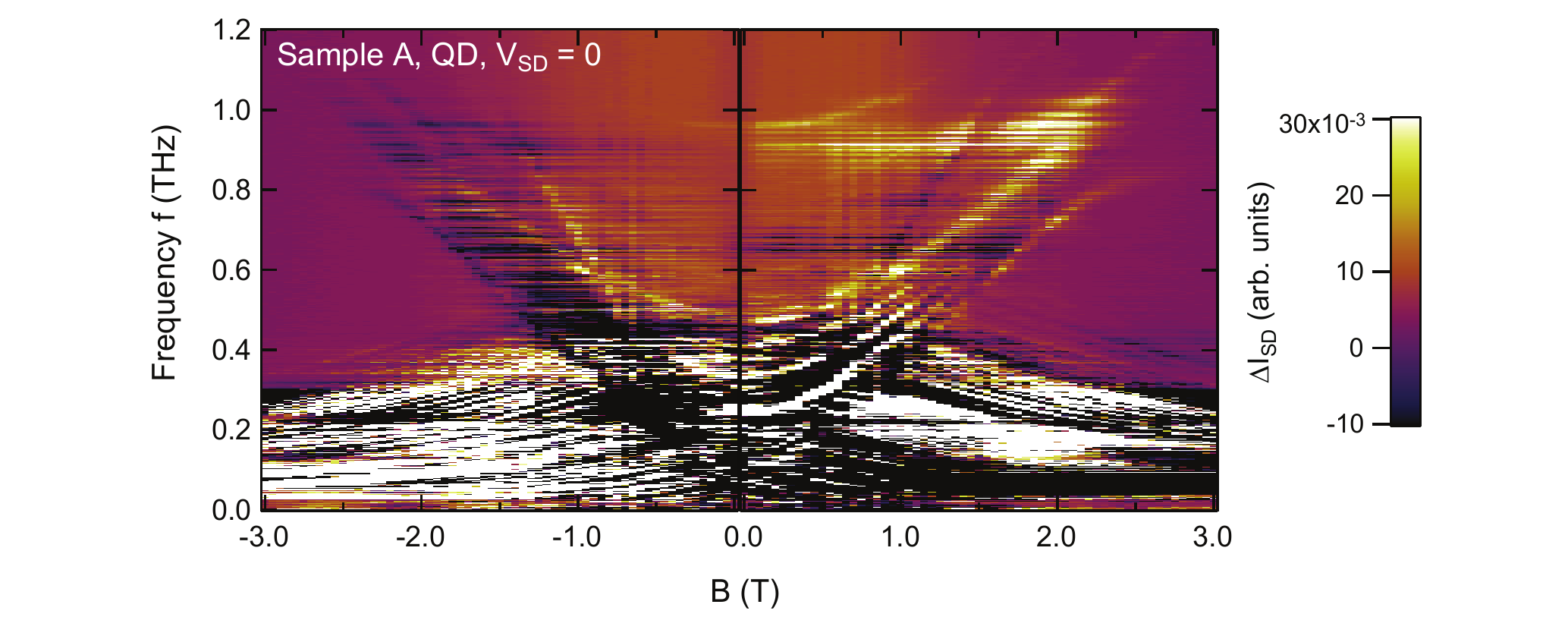} 
\caption{The same photocurrent map shown as Fig. 2(b) in the main text. For better visibility of the figure, the eye guides and arrows are removed.}
\label{figS7}
\end{figure}
\newpage
\global\long\def\bibsection{}%
\textbf{References} 
\bibliographystyle{unsrt}

%